\documentstyle[preprint,aps]{revtex}
\draft
\begin{document}
\bibliographystyle{prsty}
\preprint{RUB-TPII-6/95}
\title{Magnetic Moments of the SU(3) Octet
Baryons in the semibosonized SU(3) Nambu-Jona-Lasinio Model}

\author{H.--C. Kim
\footnote{E-mail address: kim@hadron.tp2.ruhr-uni-bochum.de},
M. V. Polyakov
\footnote{On leave of absence from Petersburg Nuclear Physics
Institute, Gatchina, St. Petersburg 188350, Russia},
A. Blotz
\footnote{Present address: Department of Physics, State University of
New York, Stony Brook, 11794, U.S.A.},
and K. Goeke
\footnote{E-mail address: goeke@hadron.tp2.ruhr-uni-bochum.de }}
\address{Institute for Theoretical Physics II, \\ P.O. Box 102148,
Ruhr-University Bochum, \\ D-44780 Bochum, Germany }
\date{August, 1995}
\maketitle
\begin{abstract}
We investigate the magnetic moments of the SU(3) octet baryons
in the framework of the $SU(3)$ semibosonized Nambu--Jona--Lasinio model.
The rotational $1/N_c$ corrections and strange quark mass in linear order
are taken into account.  We derive general relations between
magnetic moments of the SU(3) octet baryons, based on
the symmetry of our model.  These relations indicate that
higher order corrections such as $O(m_s/N_c)$ and $O(m^{2}_{s})$
are relatively small.
The magnetic moments of the octet baryons
predicted by our model are quantitatively
in a good agreement with experimental
results within about 15$\%$.
\end{abstract}

\section{Introduction}
The semibosonized Nambu-Jona-Lasinio model (NJL)
(often is called as the chiral quark soliton model)~\cite{dpp}
is very successful
in describing the static properties of the nucleon such as
the mass splitting of the nucleon
 and $\Delta$ isobar, axial constants~\cite{chretal},
magnetic moments~\cite{su2em}, electromagnetic form factors~\cite{su2em},
and so on.
 Recently, Blotz {\em et al.}~\cite{Blotzetal} and Weigel {\em et al.}
{}~\cite{Weigeletal} showed that the SU(3) version of the model explains
the correct mass splitting of the SU(3) octet and decuplet baryons.
The model could also reproduce the axial constants $g^{(0)}_{A},
g^{(3)}_{A}$ and $g^{(8)}_{A}$ with a remarkable agreement with
experiments.  In particular, the finding of the non-commutivity
of the collective operators arising from the {\em time-ordering}
enabled the model to solve the long-standing problem of
the underestimate of the axial coupling constants and nucleon magnetic
moments~\cite{chretal} in hedgehog models.

In the semibosonized NJL model, the baryon can be understood as
$N_c$ valence quarks coupled to the polarized Dirac sea
bound by a nontrivial chiral back ground field in the
Hartree approximation.  The proper quantum numbers of baryons
are obtained by the semi-classical quantization performed by
integrating over zero-mode fluctuations of the pion field around the
saddle point.  The merit of the model is to interpolate
between the naive quark model and the Skyrme model,
which enables us to study the interplay between these two models.

In the present work, we shall investigate the magnetic moments of the
SU(3) octet baryons in the SU(3) NJL model.
Since the magnetic moments of the SU(3) octet baryons are experimentally
well known, it is a good check for the model to reproduce them.
Furthermore, we shall show that the model reaches the upper limit
of the accuracy which can be attained in any model with ``
{\em hedgehog symmetry}''.

The outline of the present work is as follows: In the next section,
we briefly describe the semibosonized SU(3) NJL model
and show how to obtain the magnetic moments in the model.
In section 3, we derive the general relations between magnetic
moments of the octet baryons using the symmetry of the model
and confront them with experimental data.  We show that
subleading $O(m_s/N_c)$ and $O(m^{2}_{s})$ corrections
are relatively small, whereas $O(1/N_c$) ones are fairly large.
 In section 4, we discuss the numerical results.
We summarize the present work and draw conclusion in section 5.
\section{Formalism}
The semibosonized NJL model is described by a partition function
in Euclidean space given by the functional integral over
pseudoscalar meson and quark fields:
\begin{eqnarray}
{\cal Z} &=& \int {\cal D} \Psi {\cal D} \Psi^\dagger {\cal D}
\pi^a \exp{\left( -\int d^4x \Psi^\dagger iD \Psi\right)},
\nonumber \\
& = & \int {\cal D} \pi^a \exp{(-S_{eff}[\pi])},
\label{Eq:action}
\end{eqnarray}
where $S_{eff}$ is the effective action
\begin{equation}
S_{eff}[\pi] \;=\;-\mbox{Sp} \log{iD}.
\end{equation}
$iD$ represents the Dirac differential operator
\begin{equation}
iD \;=\; \beta (- i \rlap{/}{\partial} + \hat{m} + MU)
\end{equation}
with the pseudoscalar chiral field
\begin{equation}
U\;=\; \exp{(i\pi^a \lambda^a \gamma_5)}.
\end{equation}
$\hat{m}$ is the matrix of the current quark mass given by
\begin{equation}
\hat{m} \;=\;\mbox{diag} (m_u,m_d,m_s)
\;=\; m_0{\bf 1} \;+\; m_8 \lambda_8.
\label{Eq:mass}
\end{equation}
$\lambda^a$ represent the usual Gell-Mann matrices normalized as
$\mbox{tr}(\lambda^a\lambda^b) = 2 \delta^{ab}$.   Here, we have assumed
the isospin symmetry.  $M$ stands for the
dynamical quark mass arising from the spontaneous chiral
symmetry breaking, which is in general momentum-dependent~\cite{dp}.
We regard $M$ as a constant and employ the proper-time regularization
for convenience.  The $m_0$ and $m_8$ in eq.~(\ref{Eq:mass})
are respectively defined
by
\begin{equation}
m_0\;=\; \frac{m_u+m_d+m_s}{3},\;\;\;\;\;m_8\;=\;
\frac{m_u+m_d-2m_s}{2\sqrt{3}}.
\end{equation}
 The operator $i D$ is expressed in Euclidean space in terms of
the Euclidean time derivative $\partial_\tau$
and the Dirac one--particle hamiltonian $H(U)$
\begin{equation}
i D \; = \; \partial_\tau \; + \; H(U) + \beta\hat{m} - \beta\bar{m}{\bf 1}
\label{Eq:Dirac}
\end{equation}
with
\begin{equation}
H(U) \; = \; \frac{\vec{\alpha}\cdot \nabla}{i}
\;+\; \beta MU \;+\;\bar{m}{\bf 1}.
\label{Eq:hamil}
\end{equation}
$\bar{m}$ is defined by $(m_u+m_d)/2 = m_u = m_d$.
$\beta$ and $\vec{\alpha}$ are the well--known
Dirac hermitian matrices.  Note that the NJL model is
a low-energy effective model of QCD.  Hence, the effective action given by
eq.(~\ref{Eq:action}) can include higher order mass terms
like $\hat{m}^2 \Psi^\dagger \Psi$.  However, the coefficient
in front of $\hat{m}^2 \Psi^\dagger \Psi$ is not theoretically known
\footnote{The coefficient $\hat{m} \Psi^\dagger \Psi$ is determined by
the soft pion theorem.}.  To go beyond the linear order of mass
corrections, one should justify such a higher order term.
 Otherwise, it is meaningless to consider higher order mass corrections
in the expansion of the quark mass.  Therefore, we shall take into
account the mass corrections only up to the linear order.

 Many physical processes (semileptonic decays, electromagnetic
transitions, electromagnetic form factors, etc.) are described by
the one-current baryon matrix element:
\begin{equation}
\langle B_2 | \bar{\psi}
\Gamma \hat{O} \psi(x) | B_1 \rangle, \label{mat}
\end{equation}
where $\Gamma =
(\gamma_{\mu}, \gamma_{\mu} \gamma_5, \sigma_{\mu \nu}, \gamma_5)$
is a particular Dirac matrix depending on the physical observable and
$O$ is a $SU(3)$ flavor matrix.  For example, the matrix
element in eq.(\ref{mat}) with
\begin{equation}
\Gamma \;=\;
\gamma_{\mu},\;\;\; \hat{O} \;=\;\frac12 \lambda^3 + \frac{1}{2
\sqrt{3}} \lambda^8
\end{equation}
is relevant to the electromagnetic form
factors of the octet baryons (magnetic moments,
electromagnetic square radii, etc.).
This particular matrix element is a subject of the present paper.

One can relate the hadronic matrix element eq.(\ref{mat}) to a
correlation function:
\begin{equation}
\langle 0 | J_{B_1}(\vec{x},T) \bar{\psi}
\Gamma O \psi J^{\dagger}_{B_2}(\vec{y},0) | 0 \rangle
\label{corf}
\end{equation}
at large Euclidean time $T$ . The baryon current $J_B$ can be
constructed from $N_c$ quark fields,
\begin{equation}
J_B=\frac{1}{N_c!}\varepsilon^{i_1 \ldots i_{N_c}} \Gamma^{\alpha_1
\ldots \alpha_{N_c}}_{JJ_3II_3Y} \psi_{\alpha_1 i_1}\ldots
\psi_{\alpha_{N_c} i_{N_c}}
\end{equation}
$\alpha_1 \ldots \alpha_{N_c}$ are spin--isospin indices,
$i_1 \ldots i_{N_c}$ are color indices,
and the matrices $\Gamma^{\alpha_1 \ldots \alpha_{N_c}}_{JJ_3II_3Y}$
are chosen in such a way that the quantum numbers of the
corresponding current are equal to $JJ_3II_3Y$.  $J_B(J^{\dagger}_B)$
annihilates (creates) a baryon at large $T$.
The general
expression for the matrix elements eq.(\ref{mat}) was derived in
Ref.~\cite{ja} with linear $m_s$ corrections taken into account:
\begin{eqnarray}
\langle B_2 | \bar{\psi} \Gamma O \psi(x)
| B_1 \rangle & = & -N_c \int
d^3x e^{iqx} \int \frac{d \omega}{2 \pi}
{\rm tr} \langle x | \frac{1}{\omega
+ iH} \gamma_4 \Gamma \lambda^A \ |x \rangle \nonumber \\
& \times &
\int dR \Psi_{B_2}^{\dagger}(R) \Psi_{B_1}(R)
\frac12 tr(R^{\dagger}
\lambda^A R O)
\nonumber \\
& + & iN_c \int d^3x e^{iqx} \int
\frac{d \omega}{2 \pi} {\rm tr} \langle x |
\frac{1}{\omega + iH} \gamma_4
\lambda^A \frac{1}{\omega + iH}
\gamma_4 \Gamma \lambda^B | x \rangle \nonumber \\
& \times &
\int dR \Psi_{B_2}^{\dagger}(R)
\Psi_{B_1}(R) \frac12 tr(R^{\dagger} \lambda^A R \hat{m}) \frac12
tr(R^{\dagger} \lambda^B R O), \label{general}
\end{eqnarray}
where $q \ll
M_N$ is the momentum transfer and
$\lambda^A=(\sqrt{\frac{2}{3}} {\bf 1}, \lambda^a)$.
In eq.(\ref{general}) a regularization is not shown
for simplicity (see Ref.~\cite{kapg} for details).
$\Psi_{B}(R)$ are the rotational
wave functions of the baryon.  $\Psi_{B}(R)$ requires the corrections
due to the strange quark mass ($m_s$), since we treat the $m_s$
perturbatively.  Hence, $\Psi_{B}(R)$ can be written by
\begin{equation}
\Psi_{B}(R)\;=\; \Psi^{(8)}_{B} (R)\; +\;
c^{B}_{\bar{10}}\Psi^{(10)}_{B} (R) \;+\;
c^{B}_{27}\Psi^{(27)}_{B} (R)
\label{Eq:waveftn}
\end{equation}
with
\begin{equation}
c^{B}_{\bar{10}} \;=\; \frac{\sqrt{5}}{15}(\sigma - r_1)
\left[ \begin{array}{c}  1 \\ 0 \\ 1 \\ 0
\end{array} \right] I_2 m_s,\;\;\;\;
c^{B}_{27} \;=\; \frac{1}{75}(3\sigma + r_1 - 4r_2)
\left[ \begin{array}{c} \sqrt{6} \\ 3 \\ 2 \\  \sqrt{6}
\end{array}\right] I_2 m_s.
\label{Eq:g2}
\end{equation}
Here, $B$ denotes the SU(3) octet baryons with the spin 1/2.
The constant $\sigma$ is related to the $\pi N$ sigma term
$\Sigma\;=\;3/2 (m_u + m_d) \sigma$ and $r_i$ designates
$K_i/I_i$, where $K_i$ stands for the anomalous moments of inertia
defined in~Ref.\ \cite{Blotzetal}.
 Recently,
\cite{chretal,BloPra,ChrGor} the rotational $1/N_c$
corrections for
matrix elements of vector and axial currents
were derived, the general
expression (without any regularization) for
these corrections has a
form:
\begin{eqnarray}
\Delta_{\Omega^1}\langle B_2 | \bar{\psi} \Gamma O
\psi(x) |  B_1\rangle & = & iN_c \int d^3x e^{iqx} \int
\frac{d\omega}{2 \pi} \int \frac{d \omega^\prime}{2 \pi}
P\frac{1}{\omega - \omega'}(I^{-1})_{aa'}
\nonumber \\ & \times &
{\rm tr} \langle x | \frac{1}{\omega + iH} \lambda^{a'}
\frac{1}{\omega' +
iH}\gamma_4 \Gamma \lambda^b  | x \rangle
 \nonumber \\ & \times &
f^{abc} \int dR \Psi_{B_2}^{\dagger}(R) \Psi_{B_1}(R) \frac{1}{2}
{\rm tr} (R^{\dagger} \lambda^c R O)
 \nonumber \\ & + & N_c \int d^3x
e^{iqx} \int \frac{d \omega}{2 \pi}
{\rm tr} \langle x |\frac{1}{\omega + iH}
\lambda^{a} \frac{1}{\omega + iH}\gamma_4
\Gamma \lambda^b  | x \rangle
\nonumber \\ & \times & \int dR \Psi_{B_2}^{\dagger}(R) \frac{1}{2}
\{tr(R^{\dagger} \lambda^b R O),
\hat{\Omega}^a\}_+ \Psi_{B_1}(R).
\label{general2}
\end{eqnarray}
Where $I_{ab}$ is a matrix of the moments of inertia for the
soliton, $\hat{\Omega}^a $ is an operator of angular velocities
acting on angular variables $R$ (details can be found in
\cite{Blotzetal}).  In what follows we shall use these expressions to
calculate the magnetic moments of the SU(3) octet baryons.
\section{Magnetic moments in the Model}
Using the general expressions eq.(\ref{general}) and eq.(\ref{general2})
for  the one current baryonic matrix elements, we can express the
magnetic moments of the SU(3) octet baryons
with the $m_s$ and rotational $1/N_c$ corrections
in terms of the dynamic quantities $v_i$ depending on the
concrete dynamics of the chiral quark soliton:
\begin{eqnarray}
\mu_B & = &
 v_1 \langle B | D^{(8)}_{Q3} | B \rangle
\;+ \;\frac{v_2}{N_c} d_{ab3}
 \langle B | D^{(8)}_{Qa}\cdot \hat{J}_b | B \rangle
\nonumber \\
& + & m_s \left[(
 v_3 d_{ab3} \;+\; v_4 S_{ab3}
\;+\;v_5 F_{ab3})\cdot \langle B |
 D^{(8)}_{Q a} D^{(8)}_{8 b} | B \rangle \right] .
\label{mu}
\end{eqnarray}
Here we have introduced $SU(2)_T \times U(1)_Y$ invariant
tensors
\begin{eqnarray}
d_{abc}& = &\frac{1}{4}tr(\lambda_a
\{\lambda_b,\lambda_c \}_+),
\nonumber \\
S_{ab3}&=&\frac{1}{\sqrt3}(\delta_{a3}\delta_{b8}
+\delta_{b3}\delta_{a8}),
\nonumber \\
& & \hspace{-167pt} \mbox{and} \hfill
\nonumber \\
F_{ab3}&=&\frac{1}{\sqrt3}(\delta_{a3}
\delta_{b8}-\delta_{b3}\delta_{a8}).
\end{eqnarray}
$Q=\frac12 \lambda^3 + \frac{1}{2 \sqrt{3}}
\lambda^8$ stands for the charge operator in SU(3) flavor space.
The rotational wave functions
$|B \rangle$ are given by eq.~(\ref{Eq:waveftn}).
The dynamic quantities $v_i$ are independent of the hadrons involved.
They have a general structure like:
\begin{equation}
 \sum_{m,n} \langle n | O_1 | m \rangle
\langle m | O_2 | n \rangle f(E_n,E_m,\Lambda),
\label{spec}
\end{equation}
where $O_i$ are spin-isospin operators
changing the grand spin of states $|n \rangle$ by $0$ or $1$ and the
double sum runs over all the eigenstates of the quark hamiltonian in the
soliton field. The numerical technique for calculating such a
double sum has been developed in \cite{Blotzetal,Goetal,WaYo}.
Before we calculate the magnetic moments  numerically, let us estimate
the importance of $1/N_c$ corrections and the relative size of
subleading $O(m_s/N_c)$ corrections. To this end we employ a
dynamically independent relations between magnetic moments arising
from the ``hedgehog'' symmetry of the model.  Hyperon magnetic
moments are parametrized (in our approximation) by six parameters
($v_1,v_2,v_3, v_4, v_5$ and one parameter is contained in the
rotational wave functions).
 Looking upon them as free parameters, we
obtain the relations between the hyperon magnetic moments and the
magnetic moment of the $\Sigma^0 \Lambda$ transition :
\begin{eqnarray}
\mu_{\Sigma^0}&=&\frac{1}{2} (\mu_{\Sigma^+}+\mu_{\Sigma^-}),
\label{first}
\\ \mu_{\Lambda} &=& \frac{1}{12}(-12\mu_p-7\mu_n
+7\mu_{\Sigma^-}+
22\mu_{\Sigma^+}+3 \mu_{\Xi^-} +23 \mu_{\Xi^0})
\nonumber \\
&\times & (1+O(\frac{m_s}{N_c})+O(m_s^2))
\label{second}
\\ \mu_{\Sigma^0 \Lambda}&=&-\frac{1}{\sqrt3}(-\mu_n+ \frac14
(\mu_{\Sigma^+}+\mu_{\Sigma^-}) -\mu_{\Xi^0}+
\frac32\mu_{\Lambda})
\cdot(1+O(\frac{m_s}{N_c})+O(m_s^2)),
\label{third}
\end{eqnarray}
and one additional relation if we neglect rotational
$1/N_c$ corrections, {\em i.e.} put $v_2=0$ in eq.(\ref{mu}):
\begin{equation}
\mu_{\Xi^0}=(-3\mu_p-4\mu_n
+4\mu_{\Sigma^-}+ \mu_{\Sigma^+}+3
\mu_{\Xi^-} ) \cdot(1+O(\frac{1}{N_c})+O(m_s^2)).
\label{fourth}
\end{equation}
Note that the analogous relations between hyperon
magnetic moments was obtained by Adkins and Nappi \cite{AdkNap} but
they did not take into account mass corrections to the rotational
baryon wave functions and neglected $1/N_c$ corrections.  The
first relation  eq.(\ref{first}) is trivially fulfilled. It is
an isospin relation and so has no corrections in
both $1/N_c$ and $m_s$.  The next two relations eq.(\ref{second}) and
eq.(\ref{third}) empirically gives:
\begin{eqnarray}
 -(0.613 \pm 0.004) & = &
-(0.402 \pm 0.10) \\ & & \hspace{-7.8cm} \mbox{and} \hfill
\label{one}
\nonumber \\
-(1.61 \pm 0.08) & = & -(1.48 \pm 0.03)
\label{two}
\end{eqnarray}
 respectively, whereas
the fourth relation eq.(\ref{fourth}) gives:
\begin{equation}
-(1.250 \pm 0.015) \;=\; -(4.8 \pm 0.2) .
\end{equation}
We see that the fourth relation eq.(\ref{fourth})
in which we neglect $1/N_c$ corrections is badly reproduced by
experiment whereas the relations given by
eqs.(\ref{first},\ref{second},\ref{third}) seem to be successful.  The
explanation of this difference lies in different large $N_c$
properties of the relations.  These relations have, in
principle, corrections of order $O(1/N_c)$, $O(m_s/N_c)$ and
$O(m_s^2)$, but in (\ref{first},\ref{second},\ref{third}) all
corrections proportional to any power of $1/N_c$ are
cancelled. Hence the relations
eqs. (\ref{first},\ref{second},\ref{third}) are satisfied with the
accuracy of the order $O(m_s/N_c)$,
while the eq.(\ref{fourth}) is gratified with the accuracy of
$O(1/N_c)$.
 From these estimates one can conclude that
the corrections of order $O(1/N_c)$ to
magnetic moments numerically are large
whereas those of the order $O(m_s/N_c)$ can be relatively small.
These estimates provide us a lower limit for the
systematic errors
of computations in any ``hedgehog'' model for baryons by neglecting
the non-computed $O(m_s/N_c)$ and $O(m_s^2)$ corrections, since
any ``hedgehog'' model fulfills
eqs.~(\ref{first},\ref{second},\ref{third}) which are deviated from
the experiment by about 15~$\%$. Hence such a kind of models can not
reproduce the experimental data of magnetic moments better than
the above--mentioned limit of $15\%$.
We shall see that in the NJL soliton model
the accuracy for the hyperon magnetic moments is very close to
its upper limit.
\section{Numerical results and Discussion}
In order to calculate eq.(\ref{mu}) numerically,
we follow the well-known Kahana and Ripka method~\cite{KaRi}.

In table 1 we show the dependence of the magnetic moments of the SU(3)
octet baryons on the constituent quark mass in the chiral limit
($m_s=0$).  Both of the leading term and
the rotational $1/N_c$ corrections tend to decrease
as the constituent quark mass $M$ increases.  In this limit
the $U$-spin symmetry is not broken, so that
we have the relations
\begin{eqnarray}
\mu_{p} & = & \mu_{\Sigma^{+}},\;\;\;
\mu_{n}\;=\;\mu_{\Xi^{0}},
\nonumber \\
\mu_{\Sigma^{-}} & = & \mu_{\Xi^{-}},\;\;\;
\mu_{\Sigma^{0}}\;=\;-\mu_{\Lambda} .
\end{eqnarray}
Compared to the SU(2) results, the prediction of the SU(3) model
($m_s=0$) for the nucleon is different and seems to be better.
It is due to the fact that in our approach the nucleon possesses
the polarized hidden strangeness~\cite{BloPolGoe,kwg}.

The rotational $1/N_c$ corrections are equally important to
the other octet members as shown in Table 1.  As a result,
the total rotational $1/N_c$ corrections contribute
to the magnetic moments around $50\%$.

The symmetry breaking terms, proportional to $m_s$,
lift the $U$-spin symmetry. The $m_s$ corrections arise
from the explicit dependence of
the baryon matrix elements on the strange
quark mass (second term of eq.(\ref{general})) on the one hand,
and on the other hand come from
the solitonic rotational wave functions (details see
in Refs.~\cite{BloPolGoe,BloPra}).
The latter correction appears in each column of Table 2 and
is equally important as the former one.

It is interesting to compare the NJL model with the Skyrme model,
since these two models are closely related.
As Ref.~\cite{BloPra} already made a comparison between the NJL model and
the Skyrme model in case of the $g_A$.
 Apparently both models
have the same collective operator structures (see eq.(\ref{mu})),
but the origin of parameters $v_i$ given in eq.(\ref{mu}) is
quite distinct each other.  In the NJL model,
the coefficients $v_i$ include
the contribution from the
noncommutivity of the collective operators~\cite{chretal}
while it is absent in the Skyrme model,
since the lagrangian of the Skyrme model is local in contrast to
that of the semibosonized NJL.
 The coefficient $v_2$ in the Skyrme model comes from the
pseudoscalar mesons dominated by the induced kaon fluctuations.
 It is interesting to note that the Skyrme model needs
explicit vector mesons in addition to pseudoscalar ones~\cite{Park}
in order to achieve the same algebraic
structure of the collective hamiltonian
as it is obtained in the semibosonized NJL model with pseudoscalar
mesons alone.
Due to the introduction of vector mesons,
it is inevitable to import large numbers of parameters into
the Skyrme model.  In table 2, we compare our results with
the Skyrme model~\cite{Park}.

In Fig. 1 we show how large the predicted magnetic moments deviate from
the experimental data.  We observe that the $U$-spin symmetry is lifted
almost equidistantly.  It is due to the fact that we have only taken into
account the linear order of the $m_s$ corrections.  However,
it is not theoretically justified to consider higher $m_s$ corrections
as discussed briefly in section 2.

On the whole, the magnetic moments are in a good agreement
with the experimental data within about 15$\%$.

\section{Summary and Conclusion}
We have studied the magnetic moments of the SU(3) octet baryons
in the framework of the semibosonized SU(3) NJL model, taking into
account the rotational $1/N_c$ corrections and linear $m_s$ corrections.
The only parameter we have in the model is the constituent quark mass $M$
which is fixed to $M=420\;\mbox{MeV}$
by the mass splitting of the SU(3) baryons.
We have shown that the NJL model reproduces the magnetic
moments of the SU(3) octet baryons within about $15\;\%$.
The accuracy we have reached is more or less the upper limit which
can be attained in any model with `` {\em hedgehog symmetry}''.
\section*{Acknowledgement}
We would like to thank Christo Christov, Michal
Praszalowicz and T. Watabe for helpful discussions.
This work has partly been supported by the BMFT, the DFG and
the COSY-Project (J\"ulich).

\vfill\eject
\begin{table}[]
\caption{The dependence of the magnetic moments of the SU(3)
octet baryons on the constituent quark mass $M$ without $m_s$
corrections: $\mu (\Omega^0)$ corresponds to the leading order in the
rotational frequency while $\mu (\Omega^1)$ includes the subleading order.}
\begin{tabular}{|c||c|c|c|c|c|c|c|}
\multicolumn{1}{|c||}{\phantom{Baryonxx}}
&\multicolumn{2}{c|}{\phantom{xx}370 MeV\phantom{x}}
&\multicolumn{2}{c|}{\phantom{xx}420 MeV\phantom{x}} &
\multicolumn{2}{c|}{\phantom{xx}450 MeV\phantom{x}}
& \multicolumn{1}{c|}{\phantom{Expxx}}\\
\cline{2-7} \multicolumn{1}{|c||}{Baryon}
& \multicolumn{1}{c|}{\phantom{x}$\mu_B (\Omega^0)$\phantom{xxx}}
&\multicolumn{1}{c|}{\phantom{x}$\mu_B (\Omega^1)$\phantom{xxx}}
&\multicolumn{1}{c|}{\phantom{x}$\mu_B (\Omega^0)$\phantom{xxx}}
&\multicolumn{1}{c|}{\phantom{x}$\mu_B (\Omega^1)$\phantom{xxx}}
&\multicolumn{1}{c|}{\phantom{x}$\mu_B (\Omega^0)$\phantom{xxx}}
&\multicolumn{1}{c|}{\phantom{x}$\mu_B (\Omega^1)$\phantom{xxx}}
&\multicolumn{1}{c|}{\phantom{xx}Exp\phantom{xx}}  \\
\hline
$p$          & 1.13  & 2.57  & 1.01 & 2.27 & 0.90 & 2.11 & 2.79 \\
$n$          & -0.84 & -1.77 &-0.75 &-1.55 &-0.67 &-1.44 &-1.91 \\
$\Lambda$    & -0.42 & -0.88 &-0.38 &-0.78 &-0.34 &-0.72 &-0.61 \\
$\Sigma^{+}$ & 1.12  & 2.57  & 1.01 & 2.27 & 0.90 & 2.11 & 2.46 \\
$\Sigma^{0}$ & 0.42  & 0.88  & 0.38 & 0.78 & 0.34 & 0.72 & --   \\
$\Sigma^{-}$ & -0.28 & -0.81 &-0.25 &-0.71 &-0.22 &-0.67 & -1.16 \\
$\Xi^{0}$    & -0.84 & -1.77 &-0.75 &-1.55 &-0.67 &-1.44 &-1.25 \\
$\Xi^{-}$    & -0.28 & -0.81 &-0.25 &-0.71 &-0.22 &-0.67 &-0.65 \\
\end{tabular}
\end{table}
\begin{table}[t]
\caption{The magnetic moments of the SU(3) octet baryons
predicted by our model are compared with the evaluation from
the Skyrme model of Park and Weigel~[10].
The experimental values are taken from Ref.~[27].
The constituen quark mass
is fixed as $M = 420\;\mbox{MeV}$.  The $\mu_B (\Omega^1, m_s)$ include
subleading orders in $\Omega$ and $m_s$, which are our final values.}
\begin{tabular}{cccccc}
Baryons   &
$\mu_B (\Omega^0, m^{0}_{s})$ &
$\mu_B(\Omega^1, m^{0}_{s})$ &
$\mu_B(\Omega^1, m^{1}_{s})$ &
Park $\&$ Weigel &
$\mbox{Exp.}$ \\
\hline
$p$ & $\phantom{-}1.03$ & $\phantom{-}2.29$ &
$\phantom{-}2.39$ & $\phantom{-}2.36$
& $\phantom{-}2.79$ \\
$n$ & $-0.90$ & $-1.69$ & $-1.76$ & $-1.87$ & $-1.91$ \\
$\Lambda$ & $-0.35$ & $-0.75$ &$-0.77$
&$-0.60$ & $-0.61$ \\
$\Sigma^{+}$ & $\phantom{-}1.02$ & $\phantom{-}2.28$ &
$\phantom{-}2.42$  &$\phantom{-}2.41$
& $\phantom{-}2.46$  \\
$\Sigma^{0}$ & $\phantom{-}0.31$ & $\phantom{-}0.72$ &
$\phantom{-}0.75$ &$\phantom{-}0.66$
& $\phantom{-}$ --- \\
$\Sigma^{-}$ & $-0.40$ & $-0.85$ &$-0.92$& $-1.10$& $-1.16$ \\
$\Xi^{0}$ & $-0.74$ & $-1.54$ & $-1.64$
&$-1.96$ & $-1.25$ \\
$\Xi^{-}$ & $-0.23$ & $-0.69$ & $-0.68$
&$-0.84$ & $-0.65$ \\
$|\Sigma^0\rightarrow \Lambda|$ & $\phantom{-}0.74$
&$\phantom{-}1.42$ &$\phantom{-}1.51$
&$\phantom{-}1.74$ & $\phantom{-}1.61$ \\
\end{tabular}
\end{table}
\vfill\eject
\begin{center}
{\large Figure captions}
\end{center}

\noindent
{\bf Fig.~1}: The magnetic moments of the SU(3) octet baryons predicted by
the semibosonized NJL model.
The first column denoted by (1) shows the magnetic moments
of the SU(3) octet baryons
in case of $m_s=0\; {\rm MeV}$.  Due to the $U$-spin symmetry,
those of corresponding baryons are degenerated.
The second column denoted by (2) designates the case of
$m_s=180\; {\rm MeV}$.  The dotted lines
show the breaking of the $U$-spin symmetry
due to large $m_s$.
The third column by (3) is for the experimental data.
The constituent quark mass $M=420\;\mbox{MeV}$ is chosen for our
theoretical results.

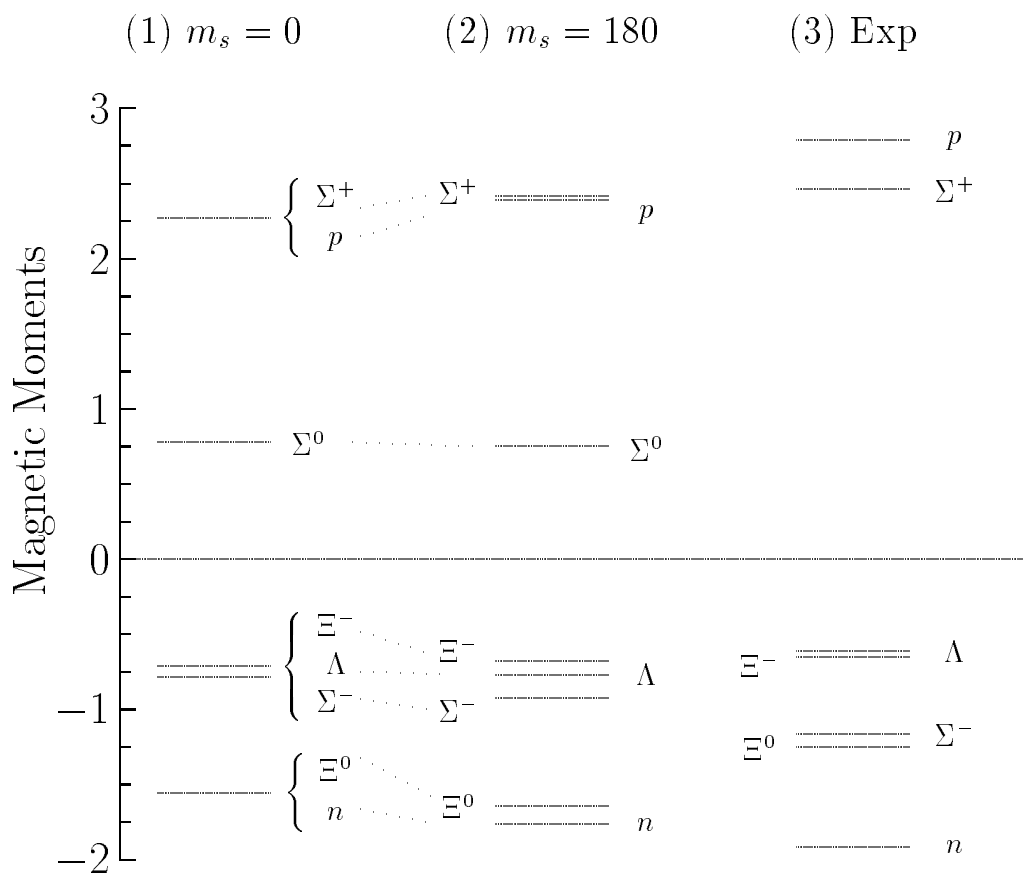
\end{document}